# Magnetic Structures and Spin States of $NdBaCo_2O_5$

Minoru Soda, Yukio Yasui, Masafumi Ito, Satoshi Iikubo and Masatoshi Sato

*Department of Physics, Division of Material Science, Nagoya University,*

*Furo-cho, Chikusa-ku, Nagoya 464-8602*

and

Kazuhisa Kakurai

*Advanced Science Research Center, Japan Atomic Energy Research Institute,*

*Tokai-mura, Naka-gun, Ibaraki 319-1195*

**Abstract**

Neutron diffraction studies have been carried out on a single crystal of oxygen-deficient perovskite $NdBaCo_2O_5$ in the temperature range of 7-370 K. There have been observed, with decreasing temperature $T$, two magnetic transitions accompanied with structural changes, to the antiferromagnetic phase at ~360 K and to the charge ordered antiferromagnetic one at ~250 K. We have studied the magnetic structures at 300 K and 7 K. At these temperatures the Co-moments have the so-called G-type antiferromagnetic structure, and Co ions are considered to be in the high spin state over the whole temperature range below $T_N$~360 K irrespective of whether the charge ordering is present or not.

corresponding author: M. Sato (e-mail: msato@b-lab.phys.nagoya-u.ac.jp)

Keywords: $NdBaCo_2O_5$, neutron diffraction, magnetic structure, spin state, charge order

# 1. Introduction

Cobalt oxides have attracted much interest, because they often exhibit the spin state change. The low spin (LS; $t_{2g}^6$; spin $S=0$) ground state of $Co^{3+}$ ions observed in many of these oxides changes to the intermediate spin (IS; $t_{2g}^5 e_g^1$; $S=1$) state or the high spin (HS; $t_{2g}^4 e_g^2$; $S=2$) state with increasing temperature $T$.[1-4] This indicates that the energy difference $\delta E$ between these spin states is small. Then, a wide variety of physical behaviors related to the spin state change may be realized by controlling $\delta E$.

In the perovskite oxides $R_{1-x}A_x CoO_{3-\delta}$ (R=Y and rare earth elements; A=Sr, Ba and Ca) a metallic and ferromagnetic phase is often induced in the region of small $\delta$ by the double exchange interaction.[5-8] For the combinations of large $Ba^{2+}$ ions and relatively small $R^{3+}$ ions, the system forms several kinds of the oxygen-deficient perovskite structure, in which the oxygen vacancies are ordered,[9-17] and a various physical behaviors are expected. However, only the restricted information is available on both the structural properties and the electronic properties.

Previously we reported results of neutron scattering studies carried out on a single crystal of $TbBaCo_2O_{5.5}$ ($\delta \sim 0.25$, for which the valence of all Co ions is +3).[14] The system has the linkage of alternating $CoO_6$ octahedra and $CoO_5$ square pyramids along the $b$-axis and exhibits several transitions with decreasing $T$, a metal to insulator transition at 340 K, to a ferromagnetic phase at 280 K and to an antiferromagnetic one at 260 K. The magnetic structure analyses have been carried out at $T=270$ K (the ferromagnetic phase) and $T=250$ K (the antiferromagnetic one), and the spin states of Co ions within the $CoO_6$ octahedra and the $CoO_5$ pyramids are clarified: The $Co^{3+}$ ions of the $CoO_6$ octahedra are in the LS state, and those of the $CoO_5$ pyramids are possibly in the IS state at both temperatures. Non-collinear nature of the magnetic structures has been proposed in both phases.

In the present work, neutron scattering studies have been carried out on a single crystal of $NdBaCo_2O_5$ ($\delta \sim 0.5$, for which the formal valence of Co ions is +2.5) to collect information on the spin state of Co ions in relation to the local structures. It belongs to the series of R'$BaCo_2O_5$ with rare earth elements R', and the structure is formed of a linkage of $CoO_5$ pyramids and Nd- and BaO-layers which are ordered along the $c$-axis as shown schematically in Fig. 1.[9-11] For R'=Y, Tb, Dy and Ho, there have been reported two transitions with decreasing $T$, an antiferromagnetic ordering of the Co-moments at around 350 K and a charge ordering at around 220 K. In YBaCo$_2$O$_5$, a simultaneous transition of the spin state has been suggested with the charge ordering.[10] However, no concrete determination of the Co spin states exists for any one of R'BaCo$_2$O$_5$ systems. Furthermore, the charge ordering transition has not been observed for NdBaCo$_2$O$_5$ in the former neutron studies.[17]

This is the first report of neutron diffraction studies carried out on a single crystal of R'BaCo$_2$O$_5$, where the antiferromagnetic transition and the charge ordering one have been observed at $T_N$~360 K and $T_{CO}$~250 K, respectively, where the spin state at the Co sites at 300 K and those of two crystallographically distinct Co sites at 7 K are determined.

## 2. Experiments

Single crystals of NdBaCo$_2$O$_5$ were grown by a floating zone (FZ) method. The crystals were checked not to have significant impurity phases by powder X-ray diffraction. The δ value of the sample was determined by the thermo gravimetric analysis (TGA) to be δ~0.51±0.02.

The electrical resistivity ρ and the susceptibility χ were measured by using edge parts of the crystal. The $T$-dependences of ρ and χ were shown in Figs. 2 and 3, respectively. The resistivity is well approximated below ~230 K by $\rho=\rho_0\exp((T_0/T)^{1/4})$ as was reported in ref. 9. An effect of the antiferromagnetic transition can be found in the χ-$T$ curve at $T_N$~360 K. We have not observed effects of the charge ordering in the curve in contrast with the case of YBaCo$_2$O$_5$.[10]

Neutron measurements were carried out by using the triple axis spectrometer TAS-2 installed at the thermal guide of JRR-3M of JAERI in Tokai. At first the crystal was oriented with the [010] (or [100]) axis vertical, where both ($h,0,l$) and ($0,k,l$) points in the reciprocal space could be reached due to the coexistence of the $a^*$- and $b^*$-domains. Measurements were also carried out for the crystal orientation with the [110] axis vertical. The 002 reflection of Pyrolytic graphite (PG) was used for the monochromator. The horizontal collimations were 17'(effective)-40'-80' (2-axis) and the neutron wavelength was ~2.359Å. Two PG filters were placed in front of the second collimater and after the sample to eliminate the higher order contamination. The sample was set in an Al-can filled with exchange He gas, which was attached to the cold head of the Displex type refrigerator for the measurements below room temperature. The can was heated in a furnace for the measurements above room temperature.

## 3. Experimental Results and Discussion

Intensities of the neutron Bragg reflections with various indices have been measured on a single crystal of NdBaCo$_2$O$_5$ in the temperature range of 7-370 K, where several superlattice reflections as well as the fundamental reflections of the unit cell with the size of ~$a_p \times a_p \times 2a_p$ have been found, where $a_p$ is the lattice parameter of the cubic perovskite cell. The $T$-dependences of the peak intensities, integrated intensities and full width at half maximum (FWHM) of ω-scans of the typical fundamental and superlattice reflections are shown in Fig. 4. The indices $h0l$ may be $0hl$ due to the existence of domains. With decreasing $T$, the superlattice peaks appear at $Q=(k'/2,k'/2,l)$ with odd $k'$ corresponding to the antiferromagnetic order at $T_N$~360 K. With further decreasing $T$, another set of superlattice peaks at $Q=(k'/2,0,l)$ with odd $k'$ appears at $T_{CO}$~250 K. These reflections are considered to have the contributions of both the nuclear and magnetic ones, as is confirmed later by detailed analyses. When the temperature of the system is lowered through $T_{CO}$, the profile widths of all reflections which exist above $T_{CO}$, begin to increase at ~$T_{CO}$ and the peak intensities are reduced with the integrated intensities being kept unchanged. The origin of the broadening is, we think, the structural distortion induced by the charge ordering at $T_{CO}$. However, the broadening does not bring about any difficulty in the analyses, because we have used only the integrated intensities.

The magnetic structures have been analyzed at 300 K and 7 K. The crystal structures have also been optimized by the simultaneous fittings, because it has been reported in refs. 9-11 that the crystal structure changes at $T_N$ and $T_{CO}$. (Space groups are P4/*mmm*, P*mmm* and P*mma* at $T>T_N$, $T_{CO}<T<T_N$ and $T<T_{CO}$, respectively.) In the analyses, we assumed that the magnitudes of all Co-moments at crystallographically equivalent sites were equal. For the magnetic form factors the average of the isotropic values of $Co^{3+}$ and $Co^{2+}$ reported in ref. 18 was used at 300 K. In the charge ordered state at 7 K, the values of $Co^{3+}$ and $Co^{2+}$ were used for Co ions with the higher and lower valences, respectively, as is stated below. The absorption- and Lorentz factor-corrections were made.

The magnetic structure at 300 K has been found to be so-called G-type antiferromagnetic one (or NaCl-type).[10] Figure 5 shows the integrated intensities of the nuclear and the magnetic reflections ($I_{obs}$) collected at 300 K against those of the model calculation ($I_{cal}$) for several reflections. The corresponding values of the magnetic reflections are also shown in Table I. Although $I_{obs}$ of several reflections with relatively strong intensity deviate from $I_{cal}$ due to the extinction, the fitting is found to be reasonable. We cannot determine the direction of the magnetic moments because of the domain distribution. The magnitude μ of the aligned moments of $Co^{2.5+}$ ions is 2.14±0.09 $\mu_B$. We presume that the spin state of Co ions is close to the HS one based on the results of the analyses at $T$=7 K shown below.

It should be noted, here, that 1/2 1/2 0 and 3/2 3/2 0 reflections, which are forbidden for the space groups reported in refs. 9-11 and for the G-type magnetic structure, appear at $T_N$~360 K with decreasing $T$. We have neglected these superlattice reflections in the analyses, because they are much weaker than the other ones (less than ~ 0.2 % of the strongest one). At this moment, we do not distinguish if the reflections exist only in the present single crystal sample or they exist in sintered samples, too. (Single crystals prepared from the molten phase may have a slightly different structure from that of sintered samples, because lattice imperfections such as the oxygen vacancies and inhomogeneity may be introduced in the course of the crystal growth.)

In the analyses at 7 K, we have assumed that Co atoms of the gray and white pyramids shown in Fig. 1 have different valences and carry different magnetic moments. Although it is not necessarily required to assume that these valences are +3 and +2, we have used here the magnetic form factors of $Co^{3+}$ and $Co^{2+}$ for these two ions, and call these ions $Co^{3+}$ and $Co^{2+}$, respectively. The magnetic structure is found to be G-type, too. Within our assumption that the magnitude of the crystallographically equivalent sites are equal, the 1/200(01/20) and 3/200(03/20) reflections are forbidden for the G-type structure if the moment directions are parallel or anti-parallel to the *b*-axis. It contradicts with the experimental observation, suggesting that the moments are not along the *b*-axis.

Then, the fitting has been carried out first for the G-type magnetic structure by assuming that the direction of all Co moments are parallel or anti-parallel to *a** and with the magnitudes of Co moments at two crystallographically distinct sites being fitting parameters. (Of course, the crystal structure has been optimized simultaneously, as is stated above.) Figure 6 shows the magnetic structures obtained by the fitting. Figure 7 shows the observed integrated intensities ($I_{obs}$) plotted against the values ($I_{cal}$) calculated in the fitting. (In Table II, observed and calculated values are compared numerically.) Again, we have neglected 1/21/20 and 3/23/20 reflections in the analyses. Although the $I_{obs}$ values of several reflections with the relatively large intensities deviate from calculated ones, $I_{cal}$ due to the extinction, the experimental results are reasonably reproduced by the fitting. However, it is not easy to determine the actual angle between the Co moment direction and the *a*-axis: The fitting is equally well if only the angle is in the region between 0 and ~50°, because there exists a correlation of the angle with the difference of the moment values between two distinct Co sites.

In the case where the moment directions are parallel or anti-parallel to [100], the magnitudes μ of the aligned moments of $Co^{3+}$ and $Co^{2+}$ are 2.68±0.04 $\mu_B$ and 2.46±0.04 $\mu_B$, respectively, while for the moment direction parallel or anti-parallel to [110], the magnitudes are 2.75±0.05 $\mu_B$ and 2.43±0.05 $\mu_B$, respectively. These values suggest that both $Co^{3+}$ and $Co^{2+}$ are in the HS state ($S$=2 for ideal $Co^{3+}$ ions; $S$=3/2 for ideal $Co^{2+}$ ions). The difference between the aligned moments of $Co^{3+}$ and $Co^{2+}$ is significantly smaller than that reported in refs. 9-11. This result does not change even when the moment direction is varied in the wide angle region between [100] and [110]. (In these previous studies, the moment direction was considered to be along the *b*-axis, because the 1/200 (or 01/20) and 3/200 (or 03/20) reflections were not observed there. If we assume that the direction of the moments is along the *b*-axis, μ values similar to those of refs. 9 and 11 are obtained.) The presently observed small difference of the moments between the crystallographically distinct sites indicates that the real charge difference between the sites may not be unity.

It is presumed that the HS state of $Co^{3+}$ and $Co^{2+}$ observed at 7 K remain up to 300 K, because the integrated intensities of the magnetic reflections do not exhibit a rapid change which suggests the spin state transition.

## 4. Conclusion

The neutron diffraction studies have been carried out on the single crystal of $NdBaCo_2O_5$, which exhibits two magnetic transitions accompanied with structural changes. With decreasing $T$, the antiferromagnetic transition takes place at $T_N \sim 360$ K, and then, the charge ordering transition occurs at $T_{CO} \sim 250$ K. The magnetic structures have been analyzed at $T=300$ K and $T=7$ K and found that the so-called G-type structure is realized in both phases. In the charge ordered phase, Co ions carry rather large spin values or they may be considered to be in the HS state. We presume that Co ions are in the HS state at 300 K, too.

Acknowledgments - Work at the JRR-3M was performed within the frame of JAERI Collaborative Research Program on Neutron Scattering. We would like to thank Dr. M. Matsuda and Mr. Y. Shimojo for the technical assistance at the spectrometer TAS-2. The work is supported by Grants-in-Aid for Scientific Research from the Japan Society for the Promotion of Science (JSPS) and by Grants-in-Aid on priority area from the Ministry of Education, Culture, Sports, Science and Technology.

References

1) R. R. Heikes, R. C. Miller and R. Mazelsky: Physica **30** (1964) 1600.
2) K. Asai, A. Yoneda, O. Yokokura, J. M. Tranquada, G. Shirane and K. Kohn: J. Phys. Soc. Jpn. **67** (1998) 290.
3) W. H. Madhusudan, K. Jagannathan, P. Ganguly and C. N. R. Rao: J. Chem. Soc. Dalton Trans. (1980) 1397.
4) G. Thornton, F. C. Morrison, S. Partington, B. C. Tofield and D. E. Williams: J. Phys. C **21** (1988) 2871.
5) M. A. Señarís-Rodríguez and J. B. Goodenough: J. Solid State Chem. **118** (1995) 323.
6) K. Yoshii, H. Abe and A. Nakamura: Mater. Res. Bull. **36** (2001) 1447.
7) C. N. R. Rao, OM Parkash, D. Bahadur, P. Ganguly and S. Nagabhushana: J. Solid State Chem. **22** (1977) 353.
8) H. Masuda, T. Fujita, T. Miyashita, M. Soda, Y. Yasui, Y. Kobayashi and M. Sato: J. Phys. Soc. Jpn. **72** (2003) 873.
9) E. Suard, F. Fauth, V. Caignaert, I. Mirebeau and G. Baidinozzi: Phys. Rev. B **61** (2000) R11871.
10) T. Vogt, P. M. Woodwark, P. Karen, B. A. Hunter, P. Henning and A. R. Moodenbaugh: Phys. Rev. Lett. **84** (2000) 2969.
11) F. Fauth, E. Suard, V. Caignaert, B. Domengès, I. Mirebeau and L. Keller: Eur. Phys. J. B **21** (2001) 163.
12) M. Respaud, C. Frontera, J. L. García-Muñoz, M. Á. G. Aranda, B. Raquet, J. M. Broto, H. Rakoto, M. Goiran, A. Llobet and J. Rodríguez-Carvajal: Phys. Rev. B **64** (2001) 214401.
13) H. Kusuya, A. Machida, Y. Moritomo, K. Kato, E. Nishibori, M. Takata, M. Sakata and A. Nakamura: J. Phys. Soc. Jpn. **70** (2001) 3577.
14) M. Soda, Y. Yasui, T. Fujita, T. Miyashita, M. Sato and K. Kakurai: J. Phys. Soc. Jpn. **72** (2003) 1729.
15) A. Maignan, C. Martin, D. Pelloquin, N. Nguyen and B. Raveau: J. Solid State Chem. **142** (1999) 247.
16) D. Akahoshi and Y. Ueda: J. Solid State Chem. **156** (2001) 355.
17) J. C. Burley, J. F. Mitchell, S. Short, D. Miller and Y. Tang: J. Solid State Chem. **170** (2003) 339.
18) P. J. Brown: *International Tables for Crystallography*, ed. by A. J. C. Wilson (Kluwer, Dordrecht, 1992) vol. C, chap. 4.

Figure captions

Fig. 1  Schematic structure of $NdBaCo_2O_5$. The white and gray $CoO_5$ pyramids are crystallographically equivalent above $T_{CO} \sim 250$ K, while they are distinct and their Co ions have different valences below $T_{CO}$.

Fig. 2  Plot of the electrical resistivity against $(1/T)^{1/4}$ ($T$ in K). The dashed line is a fit by the variable-range hopping model in the relatively low temperature region. Inset shows the $\ln\rho$-$T$ curve of the electrical resistivity.

Fig. 3  Temperature dependence of the magnetic susceptibility $\chi$ is shown. Inset shows the data around $T_N$ with the enlarged scales.

Fig. 4  Peak and integrated intensities and the widths of the $\omega$-scan profiles (full width at half maximum) of several reflections are shown against $T$.

Fig. 5  Integrated intensities of the nuclear and magnetic reflections collected at 300 K ($I_{obs}$) are plotted against those of the model calculation ($I_{cal}$) obtained with the space group P*mmm* and the G-type magnetic structure.

Fig. 6  A possible magnetic structure of $NdBaCo_2O_5$ obtained at 7 K assuming that the directions of the Co moment are parallel or anti-parallel to [100].

Fig. 7  Integrated intensities of the nuclear and the magnetic reflections collected at 7 K ($I_{obs}$) are plotted against those of the model calculations ($I_{cal}$), where the directions of the moments are assumed to be parallel or anti-parallel to [100]. The space group P*mma* and the magnetic structure shown in Fig. 6 are used.

Table I  Comparison of the observed intensities of the magnetic reflections with those calculated for G-type magnetic structure.

$T$=300 K

| h k l | $I_{obs}$ | $I_{cal}$ |
|---|---|---|
| 1/2  1/2  1 | 11476±723 | 12814 |
| 1/2  1/2  2 | 120±6 | 120 |
| 1/2  1/2  3 | 5582±272 | 5057 |
| 1/2  1/2  4 | 120±13 | 140 |
| 3/2  3/2  1 | 1455±115 | 1590 |
| 3/2  3/2  2 | 40±7 | 18 |

Table II   Comparison of the observed intensities of the superlattice reflections with those calculated for the spin structure shown in Fig. 6.

$T$=7 K

| h k l | $I_{obs}$ | $I_{cal}$ |
|---|---|---|
| 1/2 1/2 1 | 18428±1559 | 18219 |
| 1/2 1/2 2 | 340±29 | 483 |
| 1/2 1/2 3 | 9546±917 | 6400 |
| 1/2 1/2 4 | 424±39 | 497 |
| 3/2 3/2 1 | 2544±321 | 2267 |
| 3/2 3/2 2 | 156±21 | 73 |
| 1/2 0 0  /  0 1/2 0 | 45±5 | 59 |
| 3/2 0 0  /  0 3/2 0 | 27±3 | 22 |
| 1/2 0 1  /  0 1/2 1 | 280±12 | 201 |
| 1/2 0 2  /  0 1/2 2 | 683±30 | 270 |
| 1/2 0 3  /  0 1/2 3 | 284±13 | 303 |
| 3/2 0 1  /  0 3/2 1 | 143±7 | 96 |
| 3/2 0 2  /  0 3/2 2 | 254±16 | 368 |
| 3/2 0 3  /  0 3/2 3 | 103±6 | 126 |

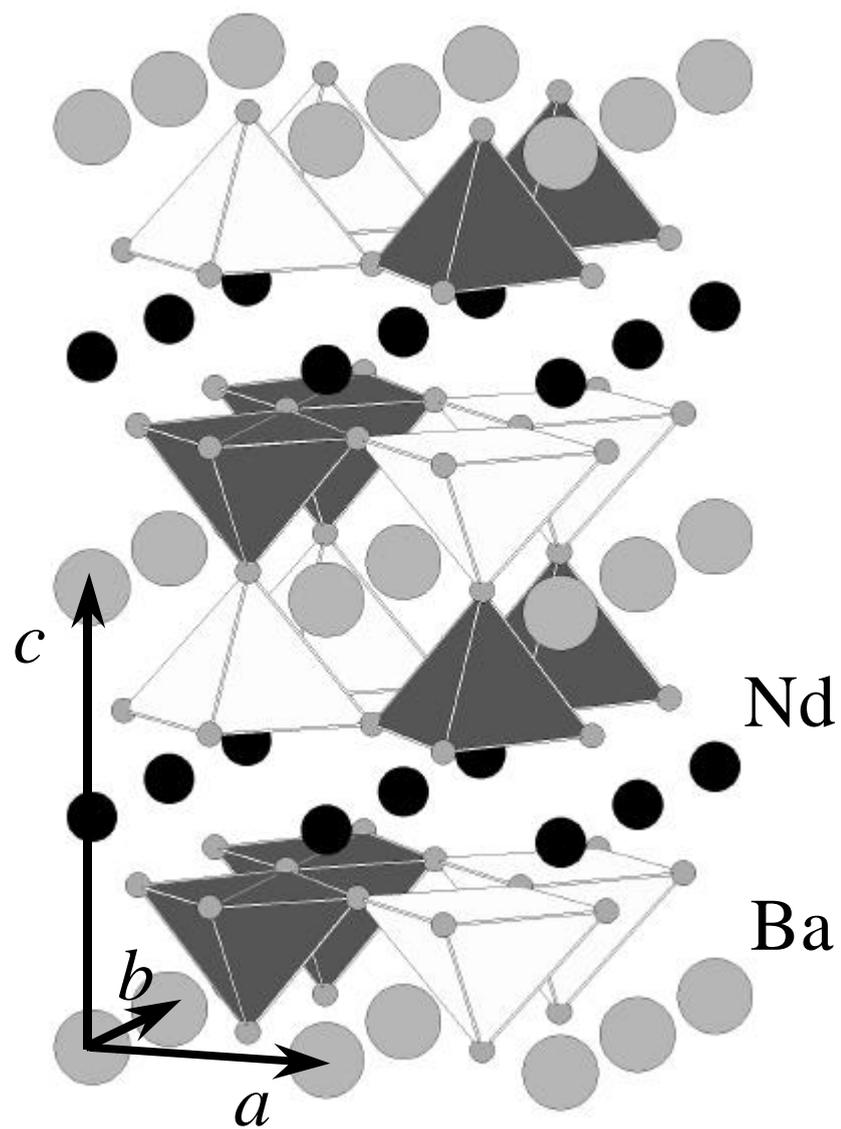

Fig. 1

M. Soda et al.

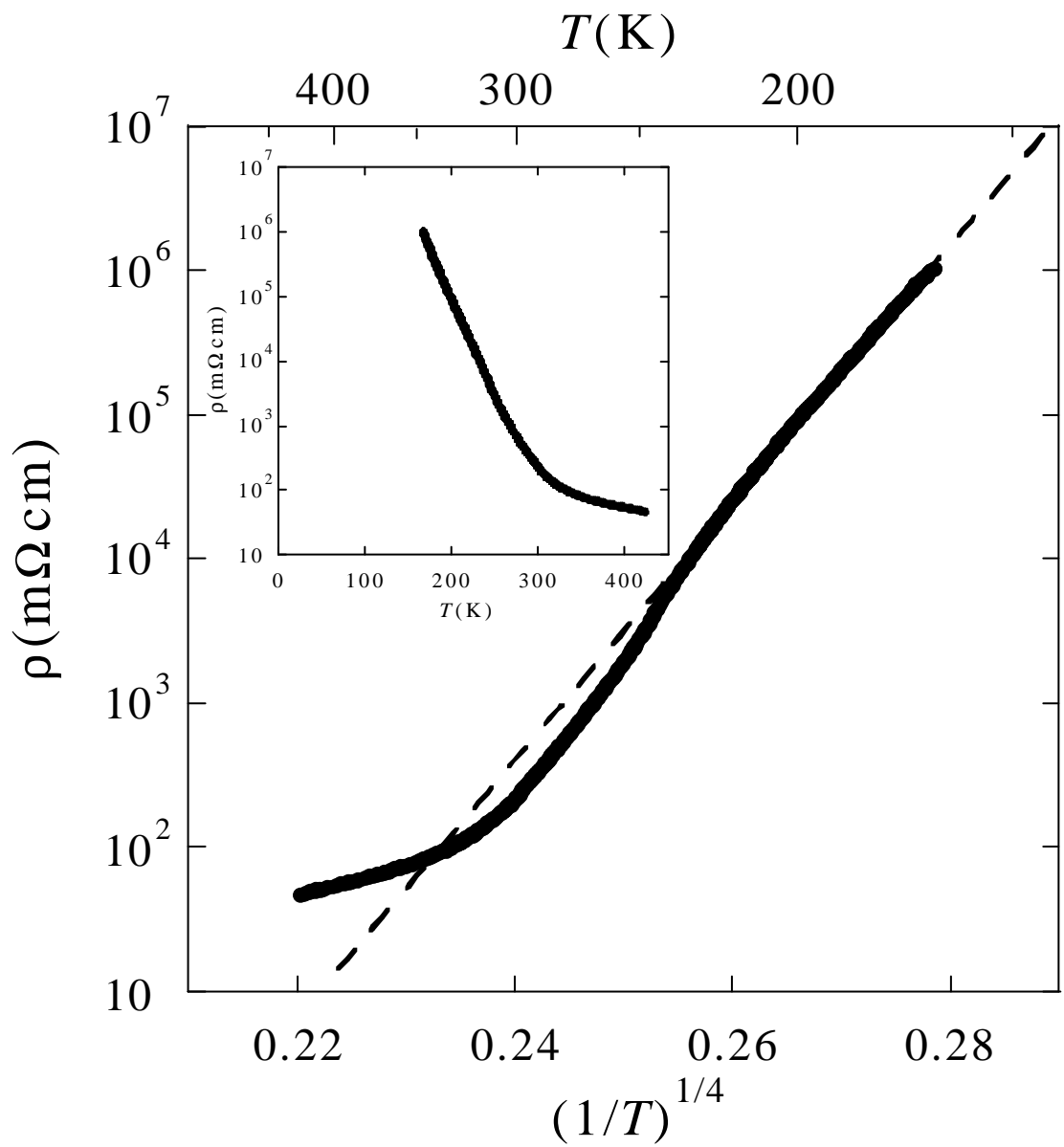

Fig. 2

M. Soda *et al.*

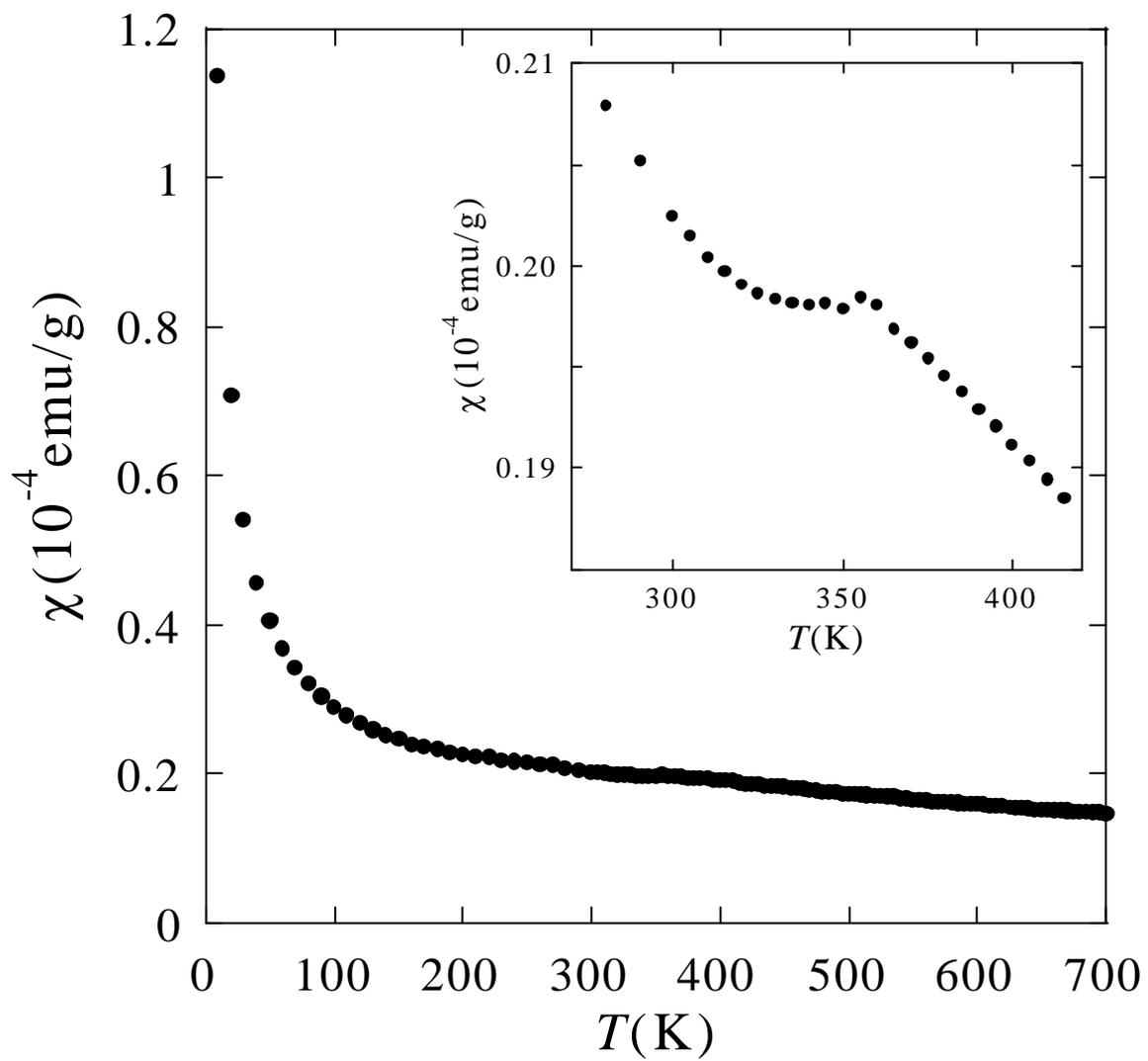

Fig. 3

M. Soda *et al.*

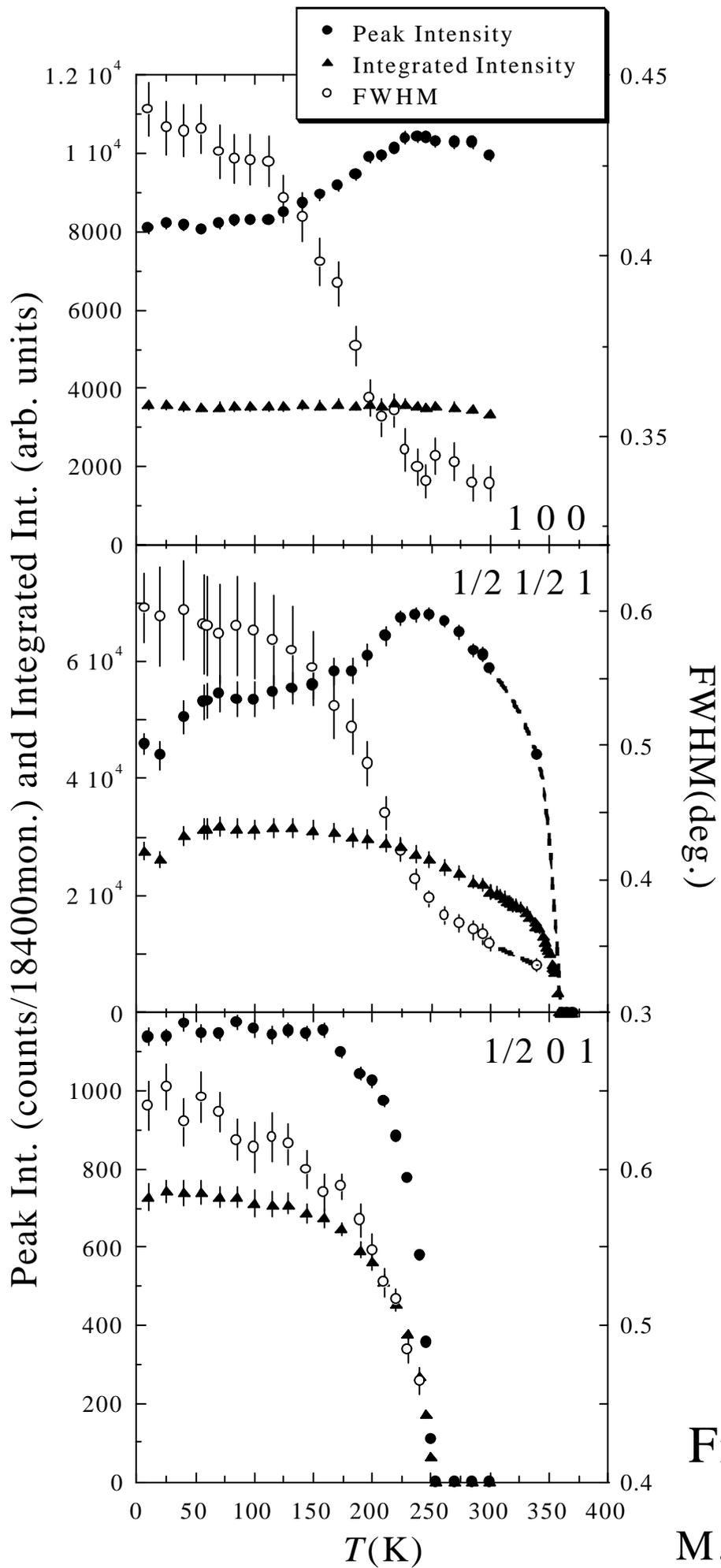

Fig. 4

M. Soda et al.

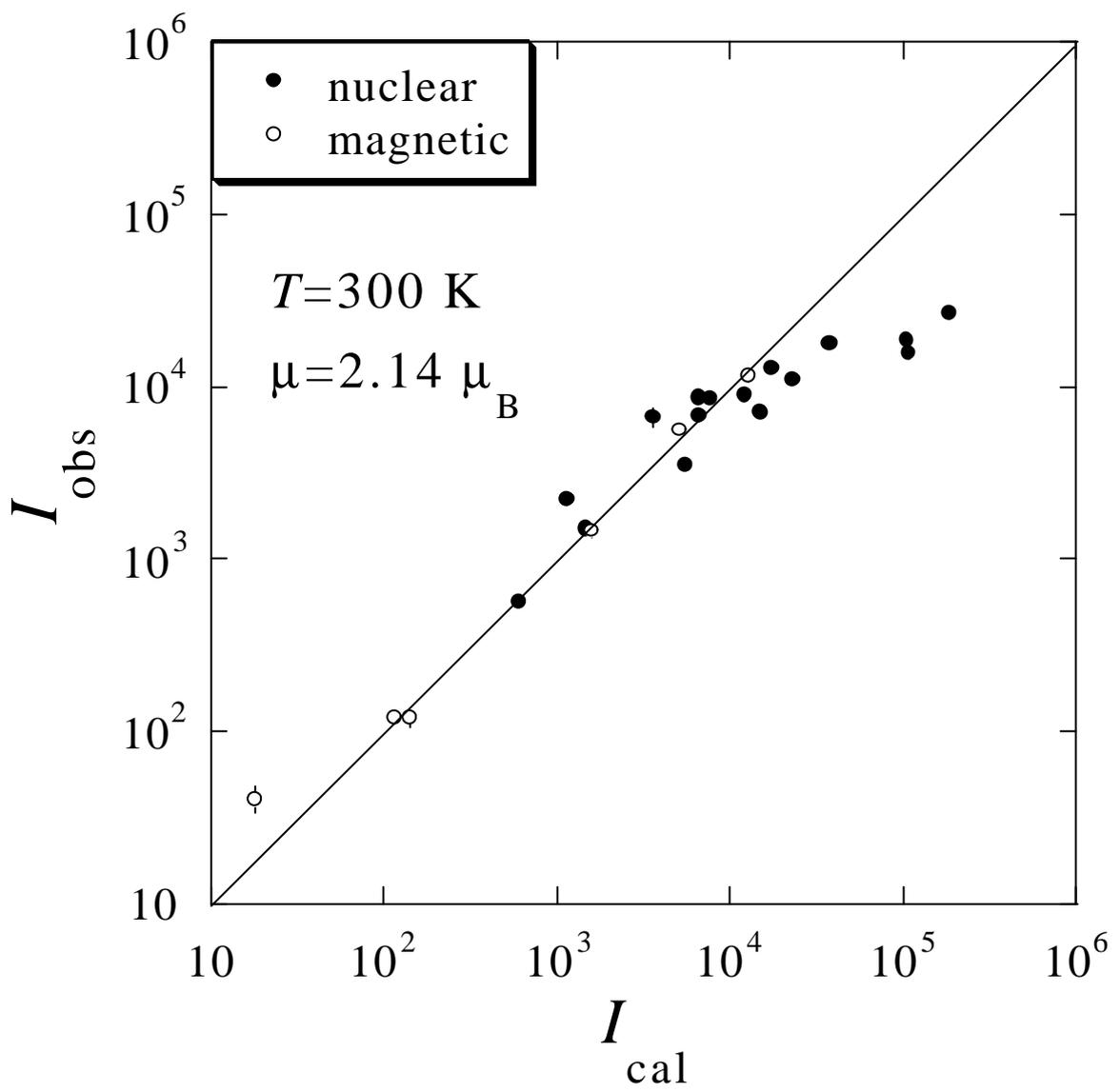

Fig. 5

M. Soda et al.

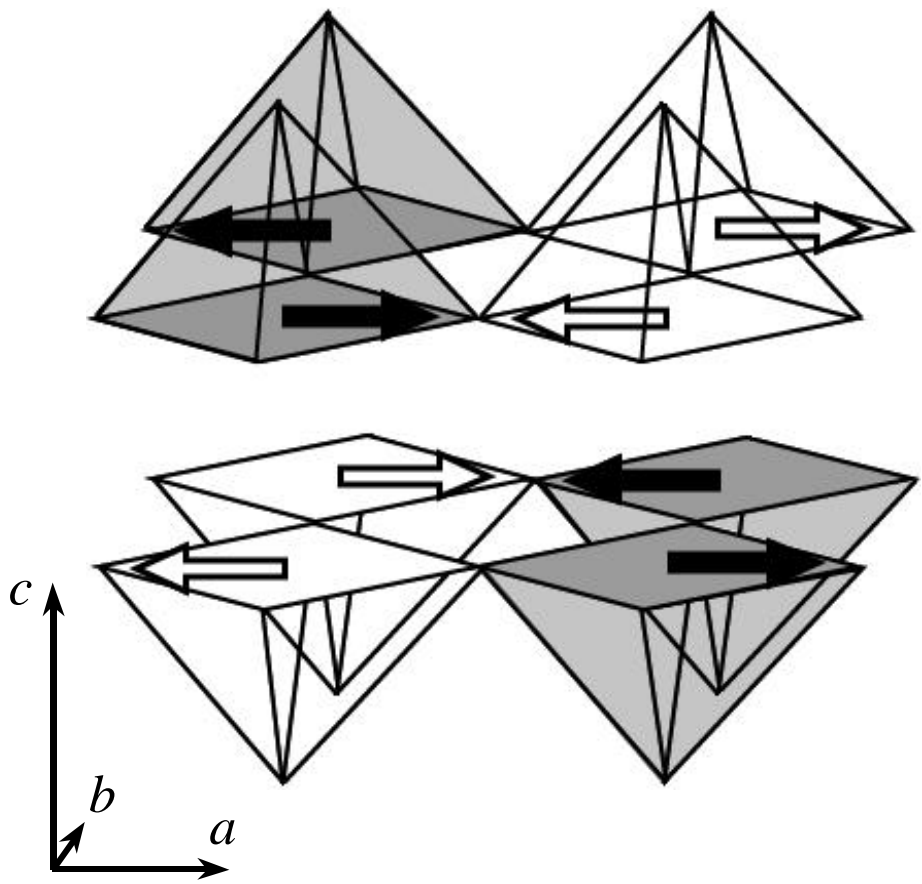

Fig. 6

M. Soda *et al.*

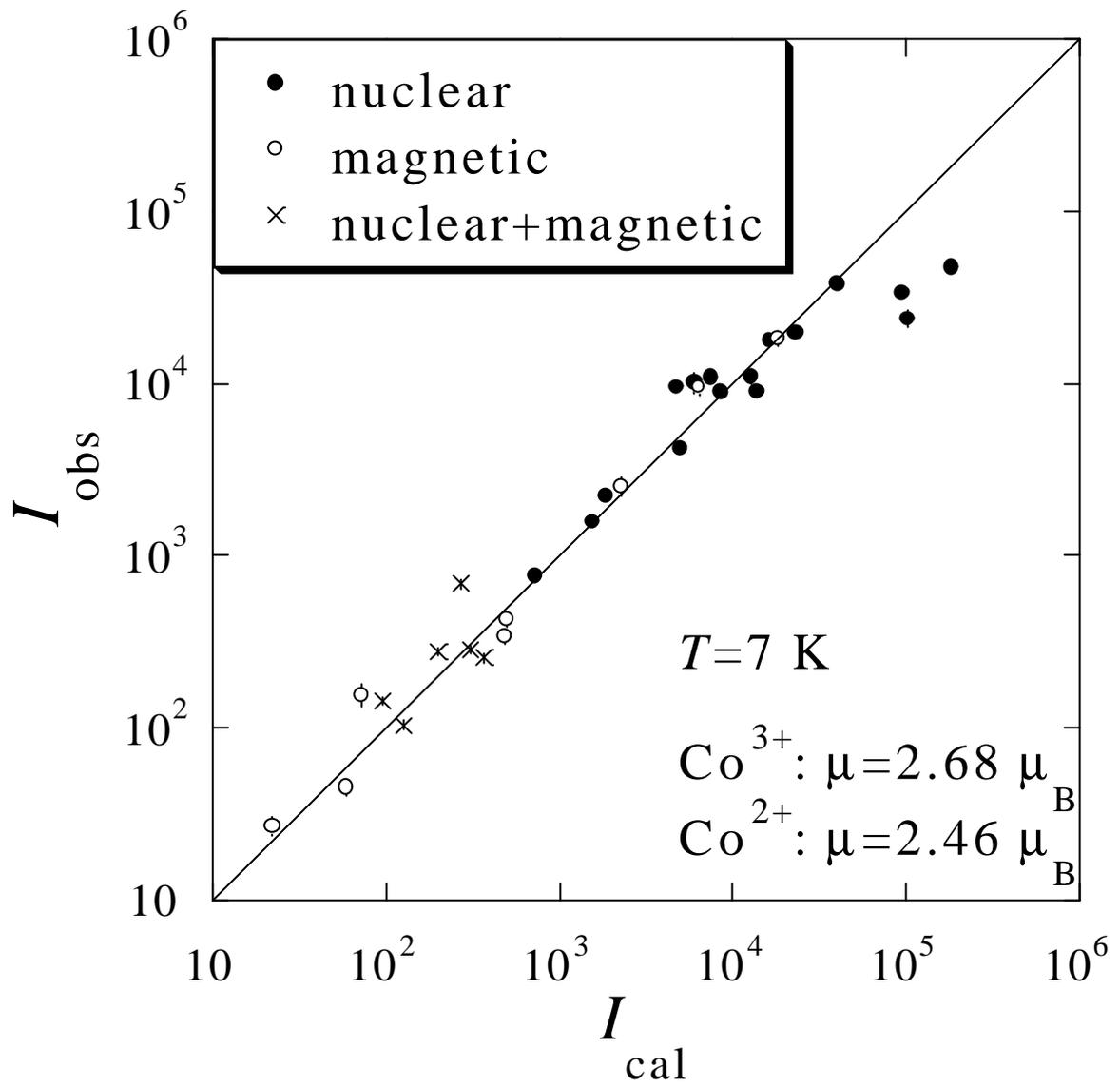

Fig. 7

M. Soda *et al.*